\documentclass[sigconf,nonacm]{acmart}

\begin{document}
\title{PromptPoseFX: Authoring Pose-Driven Visual Effects through Augmented Prompting and Interactive Refinement}
\author{Kexin Zheng}
\affiliation{%
  \institution{Hong Kong University of Science and Technology}
  \city{Hong Kong}
  \country{China}
}
\email{kzhengaj@outlook.com}

\author{Hui Ye}
\affiliation{%
  \institution{Hong Kong University of Science and Technology}
  \city{Hong Kong}
  \country{China}
}
\affiliation{%
  \institution{Hong Kong Baptist University}
  \city{Hong Kong}
  \country{China}
}
\email{huiyehy@outlook.com}

\author{Hongbo Fu}
\authornote{Corresponding author.}
\affiliation{%
  \institution{Hong Kong University of Science and Technology}
  \city{Hong Kong}
  \country{China}
}
\email{hongbofu@ust.hk}

\begin{abstract}
Human pose-driven visual effects offer a compelling way
for enhancing video expressiveness and storytelling.
Authoring such effects with traditional video editing tools is a complex and time-consuming task that requires considerable technical expertise and manual effort. While generative AI models can simplify the workflow by reducing manual scripting through language-driven generation, existing approaches suffer from imprecise alignment between visual effects and pose data, failure to preserve original video content, and limited support for fine-grained, iterative refinement.
To address these challenges, we present PromptPoseFX, an interactive system that allows creators to author pose-driven visual effects through augmented language prompts and interactive refinement. Our system leverages Large Language Models (LLMs) to translate high-level user intent into parameterized procedural graphics scripts, automatically mapping pose-tracking data to specific effect parameters, allowing precise alignment with human motion. This script-based approach manages effects as independent layers, thereby preserving the original content of the underlying video. To reduce language ambiguity, our interface offers controls across spatial, temporal, and parameter domains, enabling the precise, intuitive, and iterative authoring of pose-driven visual effects.
We evaluated PromptPoseFX through a comparative study against a text-only LLM interface, followed by an exploratory study. In the comparative study, participants rated PromptPoseFX significantly higher than the text-only baseline on measures of perceived control and precision. The exploratory study examined usability and creativity support in open-ended authoring tasks.
\end{abstract}


\begin{teaserfigure}
  \includegraphics[width=\textwidth]{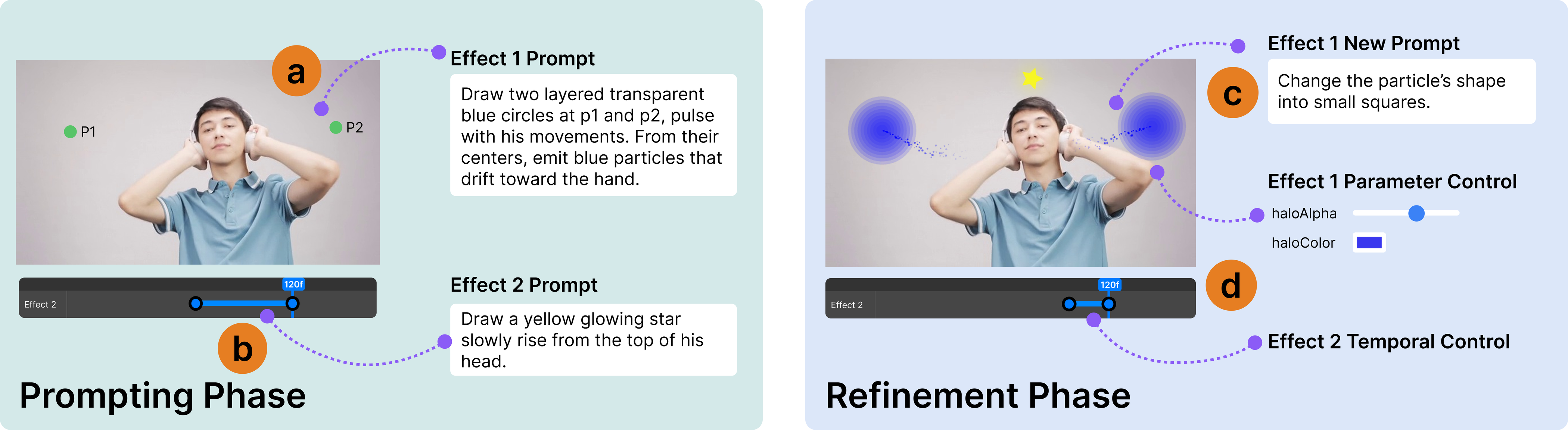}
  \caption{ PromptPoseFX enables iterative design of pose-driven effects through a hybrid interface. The workflow begins with a prompting phase, where users specify an effect by linking spatial (a) and temporal (b) information within a natural language prompt to specify the effect. In the subsequent refinement phase, users can adjust the generated result through a flexible combination of prompt editing (c) and direct manipulation of generated parameters controls and time markers in the GUI (d).}
  \Description{A two-column diagram. The left column shows  user’s text input and the created spatial and temporal control, representing the initial effect creation. The right column shows various floating UI components,sliders, a color wheel, and timeline markers, as well as new prompt written by user, illustrating the interactive editing process.}
  \label{fig:teaser}
\end{teaserfigure}

\maketitle

\section{Introduction}
Pose-driven visual effects are increasingly popular in video editing because they
synchronize visual elements with human motion to enhance expressiveness and engagement. Traditionally, creating such effects requires manual keyframing or frame-by-frame sketching, which is very complex and time-consuming. While professional tools like Adobe After Effects\cite{aftereffects} and DaVinci Resolve\cite{davinciresolve} offer motion tracking features, they mainly support attaching graphics to tracked positions.
Mapping motion data to other visual properties, whether spatial transforms like scale and rotation or non-spatial properties like color and opacity
remains labor-intensive;
creators must manually keyframe animations, use complex internal linking tools to map data to properties, or write custom scripts for sophisticated effects—all of which require significant technical expertise.

To make pose-driven visual effects more accessible, previous works have explored methods such as providing effect templates\cite{realityCanvas2023}, designing novel interfaces to map pose data to parameters\cite{interactiveBody2019,poseVEC2023}, and developing a pose-based keyframing technique\cite{poseTween2020}. Although these approaches are more novice-friendly than professional tools, they often require users to prepare graphics externally and offer limited parameter control, which limits the flexibility of creation.

Recent advances in generative AI have enabled creators to produce video visual effects directly from natural language descriptions. However,
natural language is
inherently ambiguous, making it difficult to describe spatial, temporal, and stylistic details accurately. Prior work has attempted to reduce
this ambiguity by complementing text with sketches\cite{sketchVideo2025}, control signals\cite{magicStick2025}, and motion trajectories\cite{geng2024motionprompting,lee2025generativevideomotionediting,nam2025opticalflowguidedpromptoptimization}.
But these methods are based on video generation, which
remains computationally expensive, with costs scaling for longer or higher-resolution videos.
More importantly, pose-driven visual effects require generated content to be tightly anchored to the motion in the source video. Existing
pose-conditioned generation mainly focuses on avatar performance synthesis\cite{followyourpose2024,hu2024animateanyone}, which cannot be directly adapted to augment real videos with new effects aligned to the origial human motion.
General-purpose models translate text directly into pixels without analyzing the underlying pose data, thus struggling to achieve the precise spatial alignment or compute the kinematic features (e.g., velocity, distance) necessary for high-fidelity pose-driven visual effects.

To address these limitations, we present PromptPoseFX, a novel system that enables the authoring of pose-driven visual effects by combining the expressiveness of natural language with the precision of direct manipulation. The design of PromptPoseFX is informed by an analysis of existing video editing tools and a systematic review of videos featuring pose-driven visual effects, thereby ensuring our system targets real-world creative needs.

We propose a novel approach that utilizes LLM-powered  pipeline to generate procedural graphics scripts for pose-driven visual effects. These scripts are rendered as independent layers overlaid on the original video, thus preserving the original video content for reversible editing. Instead of modifying video pixels directly, our approach generates lightweight graphical layers driven by pose-tracking data. As a result, effects can respond precisely to human motion without incurring the heavy computational cost of standard generative video models.
To make this process transparent and controllable, our LLM-based pipeline is structured into three modules (i.e., Semantic Mapper, Code Generator, Parameter Binder) that separates intent interpretation, code generation, and parameter exposure for refinement.

To further reduce  ambiguity in natural language input, our interface provides controls across spatial, temporal, and parameter domains. We allow users to create spatial references on the video directly and refer to them via text without needing to know the exact coordinates. For temporal control, we design a layered effect timeline, allowing real-time timing adjustments without regeneration. Effects appear as separate layers on the timeline for independent generation and modification. The interface also offers a set of dynamically generated parameter controls (i.e., sliders, color pickers), enabling real-time, fine-grained adjustments for iterative refinement.

We evaluate PromptPoseFX through a comparative study and an exploratory study. In the comparative study, participants rated PromptPoseFX higher on measures of perceived control than the text-only baseline. The exploratory results confirm that our system provides sufficient support for creating expressive pose-driven visual effects.

Our contributions are summarized as follows:
\begin{itemize}
    \item Key challenges of authoring pose-driven visual effects identified from an analysis of professional tools, and a proposed design space of pose-driven visual effects constructed from a systematic review of online videos.
    \item An LLM pipeline that generates procedural graphics scripts and binds their parameters to pose-tracking data, enabling non-destructive, motion-driven effects.
    \item A multimodal interface that integrates natural language with direct spatial, temporal, and parameter controls for fine-grained, iterative refinement of pose-driven visual effects.
    \item Two user studies that examine the interface's usability and compare it with a text-only baseline.
\end{itemize}

\section{Related Work}
\subsection{Authoring Pose-Driven Visual Effects}
Authoring pose-driven visual effects requires creators to align virtual graphics with human motion in video. Some prior work explored preset-based approaches, where commonly used gestures \cite{augmentedChironomia2022}, visual elements \cite{realityTalk2022}, and animation types \cite{realityCanvas2023} are incorporated into templates to streamline effect creation. While these systems reduce manual effort, they limit creative freedom by constraining customization to predefined options.

To provide more flexibility, several systems enable users to map body motion data to animation parameters. This is often achieved through interaction-driven interfaces; for example, Saquib et al. \cite{interactiveBody2019} and RealityEffects\cite{realityEffects2024} utilize direct manipulation to augment video content, whereas PoseVEC \cite{poseVEC2023} employs node-based programming to map pose data to animations. Other mapping-based approaches further analyze pose data to produce natural pose-driven effects\cite{poseTween2020,elastica2024}. However, these tools often focus on general properties like scale, position, and color, neglecting internal parameters (e.g., layer numbers for a ripple effect) and making customization of external assets difficult. Our system uses LLMs to simultaneously generate procedural graphics and bind pose-tracking data to specific parameters. This unified workflow defines content and motion together, enabling fine-grained customization without manual asset modification.

\subsection{LLM-Powered Code Generation for Creative Tasks}
The code generation capabilities of LLMs have been widely adopted across a range of creative tasks. Prior work has explored using LLMs to generate static SVG images \cite{Chat2SVG_2025_CVPR,xing2025reasonsvghybridrewardrl,mggen2025} and web UIs \cite{stylette2022,WebUI2023,misty2025}. Towards more dynamic content, Keyframer \cite{keyframer2024} and DataSway\cite{dataSway2025} synthesize motion from static SVGs. For semantic control, LogoMotion \cite{logomotion2025} employs VLMs to generate meaningful logo animations, while MapStory \cite{mapstory2025} utilizes LLMs to design editable map sequences by analyzing user-provided scripts. Unlike these works, which generate standalone code-based animations, our approach produces code for dynamic, motion-aligned graphical layers that render directly onto existing video footage.

Other work has focused on creating interactive scenes with LLMs. Recent systems support XR content creation \cite{LLMR2024, LLMER2025}, visual programming learning \cite{chatscratch2024}, controllable 2D scenes creation \cite{mographGPT2026}, and interactive procedural graphics generation\cite{spellburst2023}. Rather than creating general interactive behaviors for standalone scene elements, our system authors pose-conditioned effect logic that binds visual parameters to continuously changing body motion in video. Closer to our task, AutoVFX \cite{autoVFX2024} combines LLM-based code generation with neural scene modeling to create realistic visual effects. However, it focuses on physical simulations that interact with scene geometry and high-level positional properties. In contrast, our work generates effects that respond to fine-grained human motion by leveraging low-level joint data and user-defined spatial-temporal controls. Our approach prompts the LLM to generate code logic that interprets pose tracking data, such as joint distances and velocities, to trigger and drive visual elements, ensuring the effects reflect the continuous and dynamic nature of human movement.

\subsection{Enhancing User Control in Generative Systems}
Generative systems powered by LLMs and diffusion models can produce rich outputs from text prompts, yet text alone often lacks the precision to speficy details. Prior work has explored enhancing user control in two complementary phases: prompting and post-generation refinement.

In the prompting phase, users often struggle to formulate specific high-level instructions that align with their expected output\cite{bridgetg2024}. To address this, researchers have supplemented text with multimodal inputs to improve spatial and structural accuracy. One prominent approach utilizes spatial signals—such as edge maps, depth, and pose skeletons—to guide diffusion-based generation \cite{controlnet2023, anycontrol2024, pixelponder2025}. Other systems incorporate direct manipulation\cite{directGPT2024,pointandinstruct2024} or interactive prompt editing\cite{promptcharm2024} to offer more expressive and fine-grained control. Additionally, several works leverage vision-language models to translate sketches as a complement to text input, supporting tasks ranging from UI generation \cite{chen2025specifyuisupportingiterativeui, sketch2code2025} and code editing \cite{10.1145/3706598.3713822} to animation creation \cite{li2026sketchdynamicsexploringfreeformsketches}. However, despite these improvements in initial precision, prompting-phase controls remain primarily "one-shot" interactions. If the generated output requires minor adjustments, users must often modify the input and re-generate the entire result, leading to a tedious trial-and-error process.

To avoid costly re-generation, several works introduce interactive refinement post-generation. This includes using sliders \cite{adaptiveslider2025} or drag-based interactions \cite{dragdiffusion} to modify latent representations in diffusion models. In code-centric tasks, systems like DynaVis \cite{synavis2024}, Biscuit \cite{biscuit2024}, and Ply \cite{ply2025} dynamically generate UI widgets from natural language requests for fine-tuning specific parameters or verifing generated logic through intuitive interaction.

A few systems design controls for both phases. Narrative Motion Blocks\cite{narrativemotionblock2025} enhance user control by combining direct manipulation with text input for animation generation and providing dynamically generated sliders for precise, post-generation refinement. MoGraphGPT \cite{mographGPT2026} integrates graphical controls into prompts and auto-generates sliders from generated code for post-generation adjustment. Our system adopts this dual-phase paradigm for authoring pose-driven visual effects, which involve complex spatio-temporal alignments and stylistic properties that are difficult to specify in a text prompt. By allowing users to create spatial and temporal references during the prompting phase and providing real-time parameter controls and time markers during the refinement phase, our approach naturally combines high-level text-based intent specification with low-level interaction-based precise control.

\section{Formative Steps}
To understand the common practices in pose-driven visual effects creation, we conducted
two formative studies. First, we analyzed video editing tools commonly used for creating pose-driven visual effects, examining the potential difficulties and common interaction paradigms involved in creating such effects. Second, we examined existing videos to understand the design patterns of pose-driven visual effects.

\subsection{Analysis of Existing Video Editing Tools}
\subsubsection{Methodology}
We performed functional review\cite{functionalityReview} on five mainstream video editing tools which span the functional spectrum from high-fidelity professional suites (Adobe After Effects, DaVinci Resolve, Adobe Premiere Pro, Final Cut Pro) to template-centric creative platforms (CapCut). Our analysis focused on the interaction paradigms used to map human motion to visual assets and the user workflows required to refine these effects.
\subsubsection{Findings}
We identified the following issues in creating pose-driven visual effects with the existing commonly used tools: \\
\textbf{Flexibility vs. Accessibility in Creation Workflows}: Professional software typically allows users to create trackers and utilize that data for controlling spatial properties of effects. Yet extracting more complex information such as distances between joints usually requires programming knowledge beyond the average user. Casual tools like CapCut offer effect templates that automatically respond to body movement, making them highly accessible. But these tools do not expose the underlying tracking data to users, preventing access to even basic joint information for customization. \\
\textbf{Restricted Access to Pose Data and Mapping Mechanisms}: The mechanisms for mapping motion data to visual parameters are limited and inflexible. In professional tools, associating motion data with properties beyond position, scale, or rotation—such as using hand movement to control color or opacity—requires cumbersome workflows through graphical interfaces or custom scripting. Casual tools like CapCut provide pre-defined mappings within their templates, such as stickers that follow hand movements, but the mapping logic remains fixed inside those templates. Users cannot define new associations, such as linking the distance between hands to control color intensity. Both paradigms restrict users from freely connecting motion data to visual expression.\\
\textbf{Difficulties in Creating Customizable Graphics}: It is not easy to create customizable graphics with these tools. Professional applications offer two primary paths: using basic built-in presets, which provide limited customization, or constructing graphics from scratch using geometric primitives and hand-drawn paths, which is technically demanding and time-consuming. Casual tools like CapCut provide extensive libraries of stylized stickers and effects, making graphics easily accessible. However, these assets are usually fixed, preventing users from altering their style, behavior, or response to motion.

\subsection{Analysis of Existing Pose-driven Visual Effects}
\subsubsection{Methodology}
To understand how pose-driven visual effects are constructed in real-world videos,
we conducted a formative review of online videos containing such effects. Our goal was to identify recurring authoring patterns that could inform the design requirements of our system, with particular attention to the pose signals being used, the way they were mapped to visual behaviors, and the effect properties being controlled.

We first collected a broad pool of candidate videos from YouTube, Vimeo, and Pinterest using keyword queries such as "dance body VFX," "motion-tracked visual effect," and related platform recommendations. We then screened the results for relevance, retaining only videos that clearly contained visual effects perceptibly linked to a human performer’s body movement. To ensure diversity while keeping the dataset analytically manageable, we selected 30 videos spanning music videos, commercials, and educational content.

Since a single video could contain multiple scenes and different pose-driven effects, we segmented the 30 selected videos into 104 clips for analysis. Each clip captured a distinct scene or performance segment featuring performers and one or more pose-driven effects. We then conducted a thematic analysis of these clips to identify recurring patterns in pose signal selection, feature abstraction, mapping logic, controlled effect properties, and temporal behavior. These patterns were iteratively refined into the design space presented as follows.

\subsubsection{Design Space} \label{sec:design_space}
\textbf{Dimension 1: Pose Signal Scope}
This dimension describes the scope of pose information used as the source of control, extracted from the video subject. We identify two types of pose signals:
\begin{itemize}
    \item \textit{Single-joint Signals} are features derived from one tracked body point (e.g., the nose). These signals are commonly used when an effect must follow or attach to a specific body part.
    \item \textit{Multi-joint Signals} refer to features derived from relationships among multiple joints. These include distances, orientations, and configurations formed by multiple body points, enabling effects based on pose structure or coordination.
\end{itemize}
\textbf{Dimension 2: Feature Abstraction} This dimension characterizes how pose signals are abstracted into higher-level control features. We identify four types of features:
\begin{itemize}
\item \textit{Spatial Features} refer to raw absolute coordinate values within the video frame.
\item \textit{Kinematic Features} represent motion-related properties
derived over time, such as velocity, acceleration, and movement direction.
\item \textit{Geometric Features} describe structural relationships among joints, such as the lines, angles, or areas formed by multiple joints.
\item \textit{Semantic Features} refer to higher-level gestures or actions, such as jumping or hand clapping.
\end{itemize}
\textbf{Dimension 3: Interaction Mapping} This dimension describes the logic governing the relationship between the control signal and the visual output. We proposed the following categories in this dimension:
\begin{itemize}
    \item \textit{Direct Mapping} employs a one-to-one correspondence in which a signal influences a visual feature of the same type. For example, the position of a hand directly controls the position of a star.
    \item \textit{Indirect Mapping} connects two different types of properties. For instance, a performer's speed might determine the color intensity of a visual element.
    \item \textit{Conditional Mapping} introduces event-based logic, utilizing pose signals to trigger discrete visual events. For example, an effect could be triggered to appear when a person puts his hand on his head.
\end{itemize}
\textbf{Dimension 4: Effect Properties} This dimension categorizes the specific effect variables modified by the pose controls. It contains the following categories:
\begin{itemize}
    \item \textit{Spatial Properties} include the geometric transforms of the effect, specifically its position, rotation, and scale.
    \item \textit{Stylistic Properties} include a diverse range of aesthetic channels, such as color, transparency, and particle emission rates.
    \item \textit{State Properties} govern the logical status of an effect, such as its activation or visibility toggles.
\end{itemize}

\textbf{Dimension 5: Temporal Dependency} This characterizes how an effect's timing is controlled. There are two categories under this dimension:
\begin{itemize}
    \item \textit{Synchronous Dependency} refers to changes that are strictly tied to the subject’s motion. This includes both continuous tracking (e.g., following a hand) and discrete event triggers (e.g., a color shift occurring only when a clap is detected).
    \item \textit{Asynchronous Dependency} refers to changes that are unrelated to pose, where the effect evolves according to a creator-defined timeline. For example, a trail may transition its color over a fixed duration regardless of the subject's movement.
\end{itemize}
Through our analysis of the video corpus, we identified two key patterns in how creators apply the design space in practice. First, creators frequently adopt a composite approach, where a single visual asset leverages multiple categories across several dimensions to achieve a complex effect. Second, we found that creators frequently combine multiple independent effects, each with its own pose-logic, within a single scene. This indicates that pose-driven VFX are treated not as isolated events, but as coordinated layers that together serve the video's broader narrative.
\begin{figure*}[t]
  \centering
  \includegraphics[width=\linewidth]{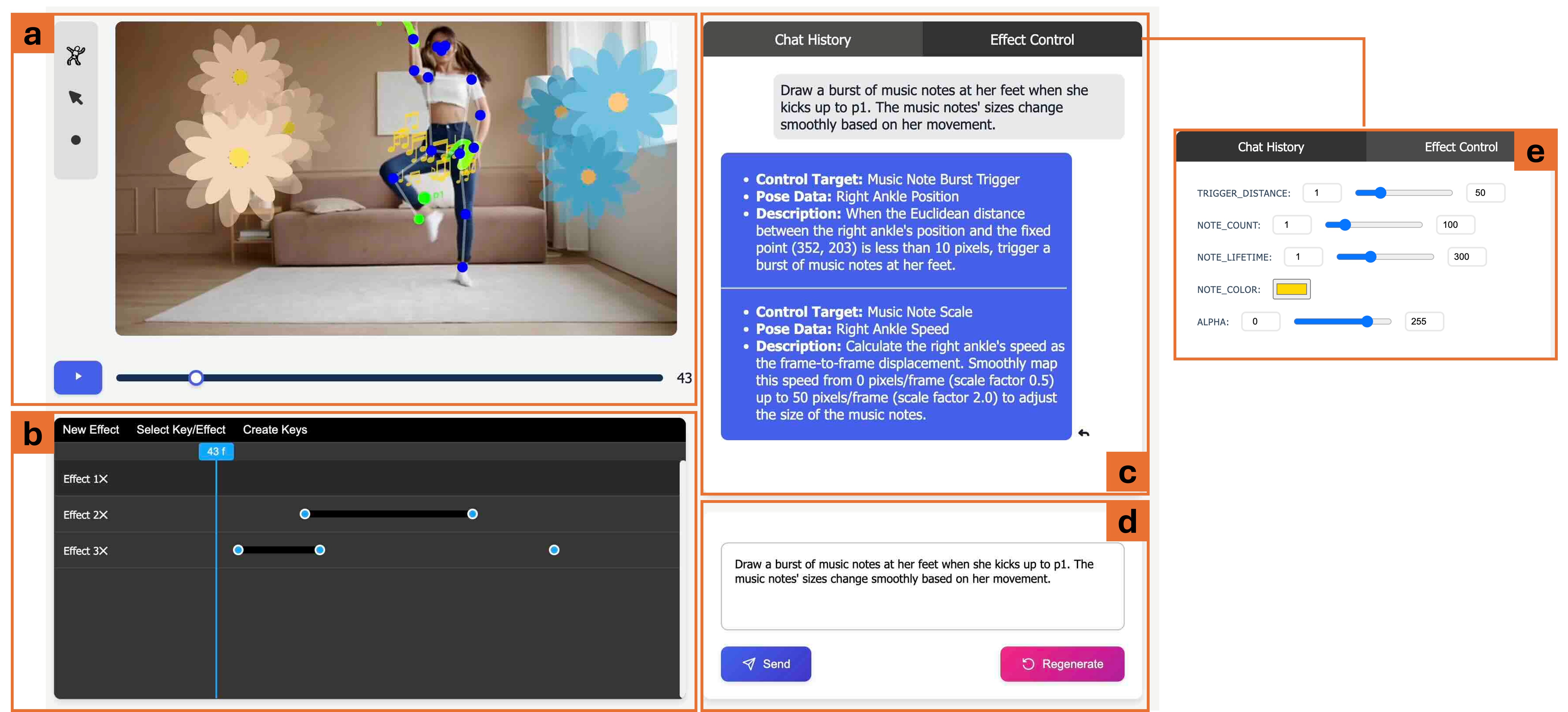}
  \caption{The user interface of PromptPoseFX. (a) \textbf{Video Canvas} allows users to view the rendered effects and add spatial control references. (b) \textbf{Layered Effect Timeline} facilitates the creation and deletion of effects, as well as the management of time-specific modifications. (c) \& (e)  \textbf{Chat History \& Effect Control Panel} displays conversation history and dynamically generated parameter control UIs. (d) \textbf{Input Box} allows users to write text prompts and to submit generation requests. }
  \Description{A screenshot of the PromptPoseFX interface featuring a video preview canvas with spatial controls, a multi-layer timeline for temporal editing, a side panel for chat history and parameter widgets, and a text input box.}
  \label{fig:systeminterface}
\end{figure*}

\section{Design Goals}
Based on the findings above, we propose three design goals to create an LLM-powered pose-driven visual effect creation system. \\
\textbf{[DG1] Lower the barrier to authoring diverse pose-driven effects. } Enable users to describe effects in natural language while supporting the wide range of pose signals, mappings, effect properties, and temporal behaviors observed in our design space. \\
\textbf{[DG2] Translate underspecified user intent into explicit \\ motion-to-effect logic. } Interpret user input to derive relevant pose features (e.g., positions, velocities, joint relationships, gestures) and map them to effect properties through appropriate direct, indirect, or conditional mappings. \\
\textbf{[DG3] Support layered and fine-grained control over effect behavior. } Allow users to independently create, adjust, and refine multiple effects across spatial, temporal, and parameter dimensions, reflecting the layered and compositional nature of pose-driven visual effects in practice.
\begin{figure*}[!t]
  \centering
  \includegraphics[width=\linewidth]{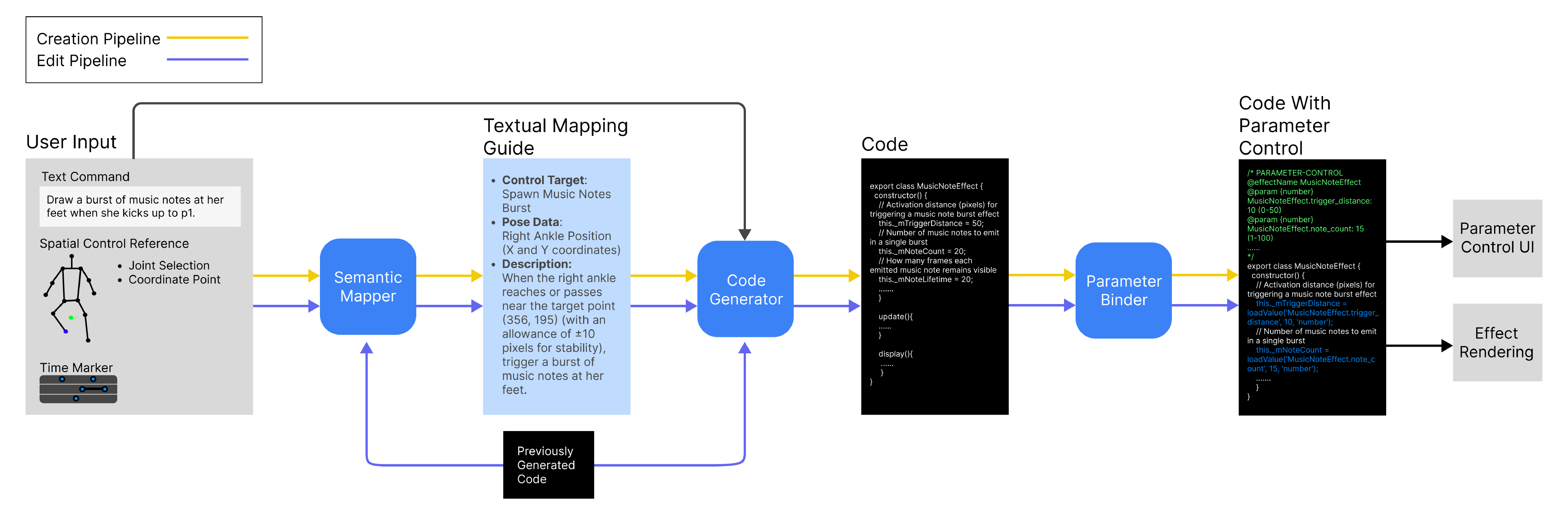}
  \caption{The LLM-powered effect generation pipeline of PromptPoseFX.}
  \Description{A flow diagram illustrating the LLM-powered pipeline in PromptPoseFX. User inputs (text, spatial, and temporal data) flow through three sequential modules—Semantic Mapper, Code Generator, and Parameter Binder—to produce a final generated effect script.}
  \label{fig:systempipeline}
\end{figure*}
\section{PromptPoseFX}
To achieve our design goals, we designed and developed PromptPoseFX, an interactive system that leverages LLMs to author pose-driven 2D visual effects. Moving beyond text input, our system supports \textbf{interaction-augmented prompting} to capture intent prior to generation, while providing tools for \textbf{interactive refinement} to fine-tune results post-generation.
\subsection{Interface}
The interface, as shown in Figure \ref{fig:systeminterface}, is organized into four primary components that provide users with intuitive and precise spatial, temporal, and parameter control for authoring pose-driven visual effects.

\subsubsection{Video Canvas}
The Video Canvas (Figure \ref{fig:systeminterface}a) displays the live video feed with rendered effects in real-time, serving as both output display and interactive surface for spatial control (DG1, DG3). A vertical toolbar provides three tools, starting from top to bottom: the Input Display Tool toggles visibility of pose skeletons and user-created reference points; the Selection Tool allows direct clicking on joints (e.g., an elbow) to bind them to effects without naming; and the Direct Spatial Coordinate Selection Tool lets users place named reference points (e.g., p1) directly on the canvas. These points can be referenced in prompts (e.g., "draw a trail from his hand to p1"), abstracting away raw coordinates. Selected joints and points are highlighted in green for immediate visual feedback.

\subsubsection{Layered Effect Timeline}
Positioned below the canvas, the Layered Effect Timeline (Figure \ref{fig:systeminterface}b) enables precise temporal control by allowing users to anchor natural language prompts to specific moments or intervals in time. Each effect occupies its own track, enabling independent generation, timing adjustment, and modification without interfering with others (DG3). To create a time-specific effect, users click "Create Keys" and then either click on a track to place a discrete time marker at that frame, or click and drag across the track to define a continuous duration between two time markers. For a time marker, users can write a prompt associated with that exact moment; for a duration, the prompt applies to the entire interval. These markers can later be dragged to new positions, and the system automatically updates the effect to occur at the new time. Selecting a keyframe or duration highlights it—orange for time markers, blue for durations—and displays its associated prompt for editing, supporting intuitive, direct manipulation of effect timing.

\subsubsection{Chat History \& Effect Control Panel}
The right panel integrates two functions (Figure \ref{fig:systeminterface}c) switchable via tabs. The Chat History tab logs all user prompts and system responses, tracing the creative process. The Effect Control tab displays real-time parameter control through dynamically generated UI elements (i.e., sliders and color pickers) for any modifiable variables identified by the backend (DG3). Adjustments made through these controls are rendered immediately on the canvas, enabling rapid, precise refinement of effect properties without reformulating text prompts.

\subsubsection{Input Box}
Users type effect descriptions in the input box (Figure \ref{fig:systeminterface}d),
This allows users to specify a wide range of pose-driven effects, motion relations, and temporal behaviors without manually constructing graphics or or animation logic (DG1).
When a time marker is selected, the box displays the prompt associated with that moment for temporal specificity. animation logic.

\subsection{LLM-Powered Effect Generation Pipeline}
The PromptPoseFX backend translates multimodal user inputs into controllable visual effects through a modular pipeline comprising three specialized LLM modules (Figure \ref{fig:systempipeline}). This modular design decomposes the complex generation task into manageable subtasks, improving system robustness. While our current implementation targets p5.js for rendering, the pipeline is easily extendable to other frameworks. Figure \ref{fig:systempipeline} illustrates the complete pipeline.

\subsubsection{Semantic Mapper}
Users often employ ambiguous and brief descriptions when prompting for visual effects\cite{Chat2SVG_2025_CVPR,dataSway2025}, making it difficult for the LLM to infer exactly what body motion should be tracked, how it should be interpreted, and which effect properties it should control.
Moreover, the internal reasoning of LLMs is typically opaque, which makes it hard for users to understand or correct mismatches between their intent and the generated result. To address these challenges, we introduce the Semantic Mapper, an LLM module that translates user intent into an explicit intermediate representation before code generation (DG2).

The Semantic Mapper takes as input the user’s natural language prompt with any spatial and temporal references created through the interface, such as selected joints, reference points, time markers, or durations. It outputs a structured pose-effect mapping specification (DG2) that serves two purposes: it provides the Code Generator with a precise specification for implementation, and it offers users textual feedback to verify whether the interpreted logic matches their intent.

The Semantic Mapper is conceptually grounded in our five-dimensional design space (Section \ref{sec:design_space}), which informs what aspects of a pose-driven effect need to be inferred from user input. To present the LLM's interpretation in a form that is intuitive for users to verify, the output is organized into three fields: what effect property is being controlled (Dimension 4), what pose data drives it (Dimensions 1 \& 2), and how they relate (Dimensions 3 \& 5).

To support both creation and refinement, the Semantic Mapper uses two templates. For new effect generation, it extracts the intended control target, pose-derived feature, mapping relation, and temporal behavior from the user input. For effect modification, it instead focuses on identifying which part of the existing effect logic should be revised. For example, given the prompt "draw a burst of music notes at her feet when she kicks up toward p1," the module may identify the target as a state property, derive a spatial relation between the foot and reference point p1, and encode a conditional trigger that activates the effect when the kicking motion satisfies the specified condition (Figure \ref{fig:systeminterface}). By externalizing this logic in a readable form, the Semantic Mapper improves both controllability and explainability before code is generated.

\subsubsection{Code Generator}
The Code Generator produces executable p5.js scripts based on the user input
and the mapping specification from the Semantic Mapper.
The generator is provided with a predefined global function for accessing joint positions. Inspired by Spellburst\cite{spellburst2023}, the module generates concise code comments that document its reasoning process to improve final code quality.

To support independent effect management (DG3), each visual effect is generated as an independent JavaScript class and linked to a track in the Layered Effect Timeline (Figure \ref{fig:systeminterface}b). The generation process follows a code template (see Appendix \ref{codeTemplate}) that separates parameter initialization from rendering logic, providing a clear structure for subsequent parameter binding and a consistent interface for executing rendering functions across all effects. We employ Retrieval-Augmented Generation (RAG) to improve code quality: the system performs semantic search over a dataset of effect description-code pairs, selecting the most similar example to guide generation.

\subsubsection{Parameter Binder}
To support iterative refinement, our system identifies and exposes controllable parameters from the generated code. We categorized visual effect parameters into two types: 1) dynamic values that are continuously updated by effect logic and 2) static values that can be freely modified without affecting the logic. For example, in an effect where multiple stars are emitted from a person's right hand, the emitter's position is a dynamic value that updates with the hand's movement while the star's color and emission rate are static values. The Parameter Binder is designed to automatically identify static parameters and make them accessible to the frontend UI (DG3). Given the generated code as input, it searches for static parameters that are suitable for user controls based on the code logic. For each identified parameter, the module refactors the code with a variable that can be updated externally by the frontend and writes a formatted comment containing the parameter's name, data type, default value, and a suggested range(See Appendix \ref{parameteroutput}). The frontend then parses this formatted comment to dynamically render the appropriate UI controls.

\subsection{Implementation}
PromptPoseFX is a web application built with p5.js as the graphics library. It executes dynamically generated effect scripts in a dedicated iframe to ensure real-time feedback. Scripts are stored as ES6 modules and managed by a master p5.js sketch.

The system maps spatial references, joint selections, and time markers to textual representations, combining them with text into a unified JSON structure for LLM reasoning.

Video is processed via MediaPipe \cite{mediaPipe2025} for 2D pose tracking. Raw landmarks are smoothed using the One Euro Filter\cite{oneEuroFilter2012} to reduce jitter. Results are stored in a frame-indexed JSON. To improve LLM accuracy, a predefined access function maps semantic joint names (e.g., "left shoulder") to their specific indices, providing the generated code with reliable positional data.

The backend leverages OpenAI's API\cite{chatGPT2025}. The Semantic Mapper and Code Generator use the o3-mini reasoning model, while the Parameter Binder adopts GPT-4o-mini for faster generation.

\section{Comparative Study}
We conducted a within-subjects comparative study to evaluate the performance of PromptPoseFX against a text-only baseline.

\subsection{Participants} We recruited 12 participants (U1–U12, 11 females, 1 male aged 25–32) through personal and university networks. All participants reported strong interests in creating pose-driven visual video effects. They came from diverse backgrounds, including computer science (U1, U9), applied math (U2), media (U3), public relations (U4), music (U5), freshwater ecology (U6), social work (U7), design (U8), psychology (U10), business (U11), and russian language (U12).
On a 5-point self-rating scale (1 = no experience, 5 = expert), participants reported varying levels of experience with commercial video-editing software like Adobe Premiere and CapCut, including three advanced (U7–U9), four intermediate (U1, U3–5), three beginner (U2, U11–12), and two with no prior experience (U6, U10). All participants had prior experience with LLMs (e.g., ChatGPT, DeepSeek, Doubao), with 9 (U1-2, U4, U7-U12) reporting frequent daily use and 8 (U1, U4-5, U7-9, U11-12) having used generative AI for visual content creation.
\subsection{Study Design and Task}
\begin{figure}[ht]
    \centering
    \includegraphics[width=\linewidth]{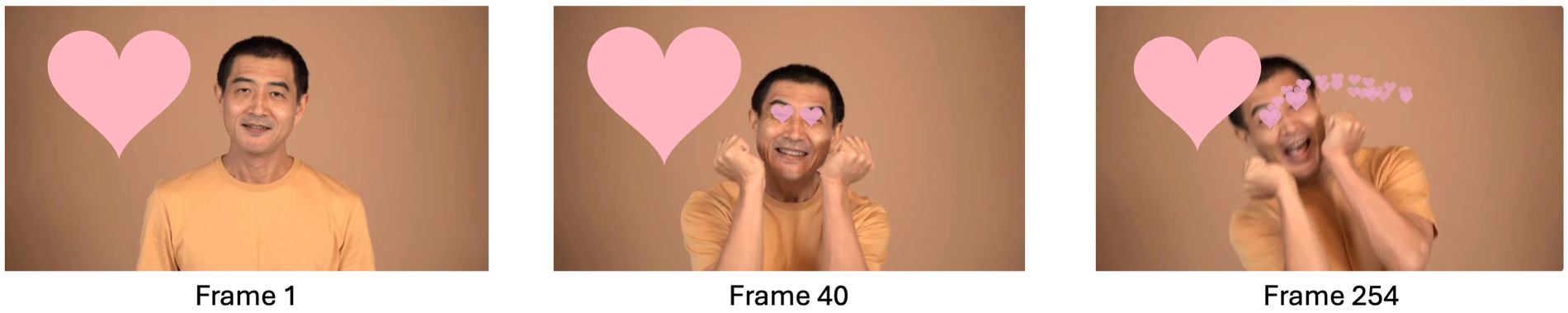}
    \caption{The target video used in the comparative study task.}
    \label{fig:comparativeVid}
    \Description{A video sequence demonstrating two generative effects: Effect 1 shows heart overlays on a man's eyes triggered by hand gestures, followed by a heart particle stream after frame 245. Effect 2 shows a heart in the top-left corner whose pulse rate scales with the man's movement speed.}
\end{figure}
We employed a within-subjects design to compare our system against a text-only baseline system. The baseline retained the full backend generation pipeline of our system but removed all additional GUI features, resembling existing text-based video generation/editing tools like Runway AI and Pika Labs.

Participants were asked to replicate two predefined pose-driven visual effects on a provided video clip (Figure \ref{fig:comparativeVid}): Effect 1 required two hearts appear at the man's eyes when he raised his hands near his face, followed by a continuous emission of small hearts from frame 245 to the end. Effect 2 required a separate heart in the top-left corner to beat faster when the person moved quickly and slower when he moved slowly.

Participants implemented both effects within a single session for each system. The system order was counterbalanced to mitigate learning effects.

\subsection{Procedure}
Both systems were deployed as web applications on a standard laptop. Eleven participants (U2–U12) participated in person; one (U1) attended remotely via Google Remote Desktop. All used a standard keyboard and mouse.

The study begins with collecting participant information with a demographic form. After that, participants received a 10-minute tutorial covering key features of both systems, followed by 5 minutes of free exploration and questions. Participants were shown a reference video of the target effects, which they could review throughout.
They were required to complete the tasks within 30 minutes for each system. If a participant failed to complete the task within 30 minutes, the attempt was recorded as failure. Task completion was verified by the experimenter against visual similarity. After using both systems, participants completed a questionnaire with NASA-TLX and Likert-scale questions(See Appendix \ref{surveyitem}) measuring perceived precision, ease of control, and usefulness of specific interaction features, followed by a 5-minute semi-structured interview.

\subsection{Results}
\subsubsection{Overview}
\begin{figure}
    \centering
    \includegraphics[width=\linewidth]{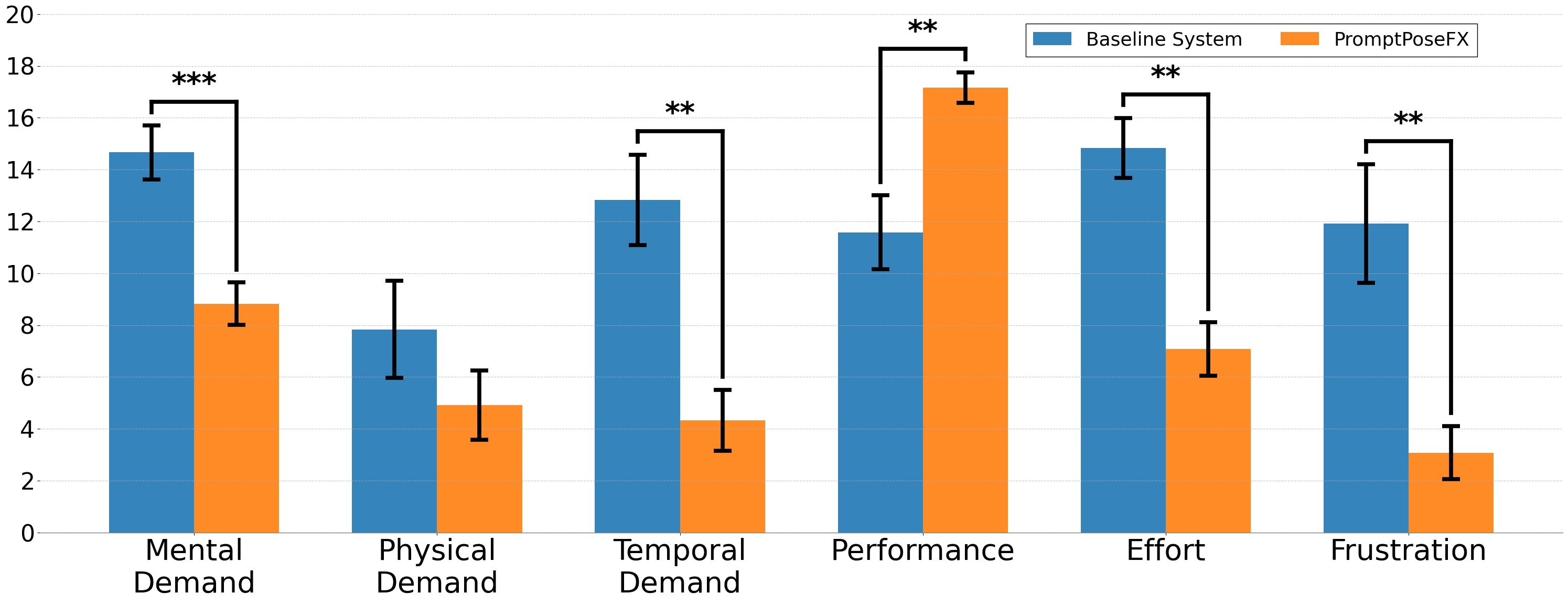}
    \caption{The NASA-TLX results. Statistical significance is denoted by $**$ ($p < .01$), and $***$ ($p < .001$).}
    \Description{A bar chart comparing NASA-TLX scores on a scale of 0–20 for the baseline (blue) and PromptPoseFX (orange). PromptPoseFX has lower scores for Mental Demand, Physical Demand, Temporal Demand, Effort, and Frustration, and a higher score for Performance. Asterisks mark significant differences based on Wilcoxon signed-rank tests ($p < .01$ and $p < .001$).}
    \label{fig:nasaTLX}
\end{figure}
All 12 participants successfully completed both effects with PromptPoseFX within the time limit. With the baseline system, three participants (U6, U7, U10) failed to complete the task within the given time.
As shown in Figure \ref{fig:nasaTLX}, participants reported significantly lower workload with PromptPoseFX across most dimensions
based on the Wilcoxon signed-rank test results. Participants expressed frustration with the baseline due to the absence of controls and inability to express desired edits through language alone (7/12). In contrast, all participants agreed that PromptPoseFX's additional functionalities enabled more precise and easier replication of target effects.

\begin{figure}
    \centering
    \includegraphics[width=\linewidth]{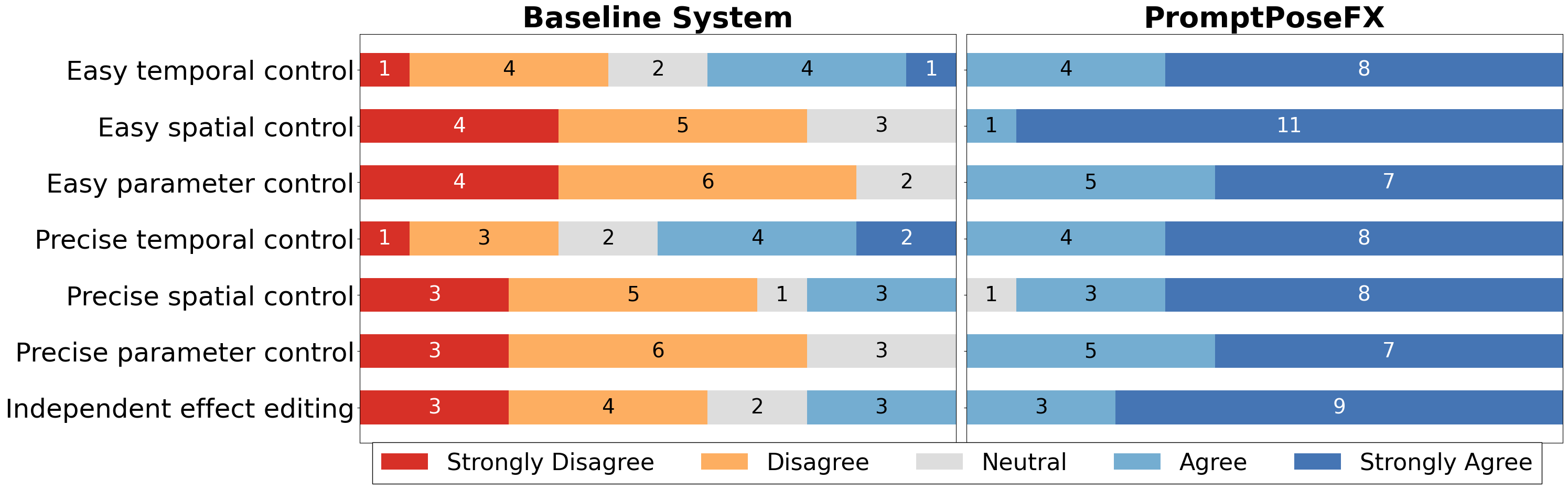}
    \caption{Subjective rating distribution on \textit{PromptPoseFX} and Baseline system.}
    \Description{A horizontal stacked bar chart showing that PromptPoseFX consistently outperforms the Baseline interface. The positive rating segments (Agree and Strongly Agree) are significantly larger for PromptPoseFX across all categories: spatial, temporal, parameter control, and independent editing.}
    \label{fig:likert}
\end{figure}

\subsubsection{Spatial Control}
Participants rated PromptPoseFX significantly higher for precise spatial control (PromptPoseFX: Mean = 4.58, SD = 0.67 ; Baseline: Mean = 2.33, SD = 1.15; p < 0.001) and for easy spatial control
(PromptPoseFX: Mean = 4.92, SD = 0.29; Baseline: Mean = 1.92, SD = 0.79; p < 0.001) compared to the baseline. When placing the large heart for Effect 2, participants often struggled to describe positions using vague terms like "upper-left corner", "left side of the person's head" (12/12), or attempted to manually estimate pixel coordinates based on the frame size (5/12). These methods lacked precise guidance, which often led to undesired results that required further adjustment. In comparison, PromptPoseFX allowed participants to select the target position via clicking, which provided a more intuitive interaction for participants to position visual effects.

\subsubsection{Temporal Control}
PromptPoseFX also outperformed baseline for temporal precision (PromptPoseFX: Mean = 4.67, SD = 0.49; Baseline: Mean = 3.25, SD = 1.29; p < 0.01) and easy temporal control
(PromptPoseFX: Mean = 4.67, SD = 0.49; Baseline: Mean = 3.0, SD = 1.21; p < 0.01). Participants usually explicitly mention the timestamp in their prompt (e.g. draw two small hearts on the person's eyes after frame 37) when using the baseline system. Though some participants (3/12) explicitly mentioned that they preceived no significant difference when writing prompts to specify time-related operations with both system, U10 noted that associating the prompts with the UIs allowed them to modify each prompt more easily than when using the text-only baseline.
Some user(3/12) also mentioned that the interaction-based timing adjustment help them save time for writing new prompts to specify time-related modifications.

\subsubsection{Parameter Control}
Participants found generated parameter control UIs is simpler (PromptPoseFX: Mean = 4.58, SD = 0.51; Baseline: Mean = 1.83, SD = 0.72; p < 0.001) and more precise (PromptPoseFX: Mean = 4.58, SD = 0.51; Baseline: Mean = 2.00, SD = 0.74; p < 0.001) to use. U3 and U6 valued real-time feedback of the paramter control; U6 noted that adjusting attributes like color of the hearts in the baseline was ambiguous and time-consuming. Participants reported difficulty in precisely modifying effect properties such as size, opacity, and emission rate, frequently relying on vague descriptions (e.g., "make the heart look darker"). Most of them went through several rounds of trial and error to achieve the desired result(10/12). In comparison, they reported that the generated parameter controls in our system simplified these edits. However, two participants (U2, U8) mentioned that generated controls sometimes lacked desired parameters, requiring a return to text prompts.

\subsubsection{Layered Editing}
Participants rated layered editing highly for independent effect modification (PromptPoseFX: Mean = 4.75, SD = 0.45; Baseline: Mean = 2.42, SD = 1.16; p < 0.01). In the baseline system, participants frequently encountered issues where new prompts accidentally overwrote or corrupted previously established effects. To avoid such issues, participants tried explicitly requesting the system to preserve previous steps using phrases like "keep the previous effect unchanged" (6/12) or manually specifying the editing targets (e.g., "the heart at the top left corner") (11/12). Another observed method was to describe every details of the target effects in one prompt (3/12). Some users reported feeling stressed about losing progress with the baseline, while all participants agreed that layered editing reduced errors and improved organization.

\subsubsection{Textual Mapping Guide}
\begin{figure}
    \centering
    \includegraphics[width=\linewidth]{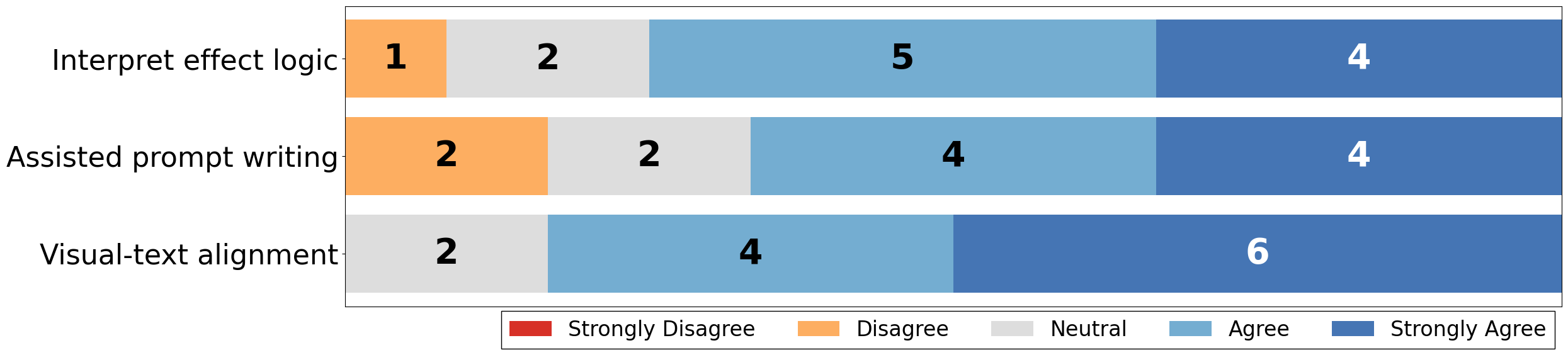}
    \caption{User response related to textual guide.}
    \Description{A horizontal stacked bar chart showing user ratings for the textual mapping guide. For all three categories:alignment, logic understanding, and prompt assistance, the responses are mostly positive.}
  \label{fig:textual-mapping-eval}
    \label{fig:textguide}
\end{figure}

Most participants found the Semantic Mapper's textual mapping guide useful (Figure~\ref{fig:textguide}). Two participants with less technical experience (2/12) found it too complex. For example, when generating a guide for the 'heartbeat' effect, the system wrote an equation describing how to use nose movement speed to control the scale of the heart. In contrast, participants with programming or advanced editing experience found the guide useful for inspecting effect logic and guiding further edits.

\section{Exploratory Study}
Eight participants from the comparative study (U1–U8) took part in the exploratory study, which focused on evaluating the usability and creativity support of PromptPoseFX.
\subsection{Study Design and Task
}
We asked the participants to complete open-ended creative task in a freeform manner.
Participants selected their own videos and were encouraged to freely create pose-driven 2D visual effects with no restrictions on style or complexity. We aimed to gather qualitative insights into creative workflows and quantitative measures of usability and creativity support.
\subsection{Procedure}
Participants used only PromptPoseFX with their self-selected videos. They were encouraged to think aloud during the process. Upon completion, participants completed a questionnaire containing the System Usability Scale (SUS) and the Creativity Support Index (CSI). The "Collaboration" dimension was removed from CSI as our system does not include collaboration features. The session concluded with a 10-minute in-depth interview focusing on overall user experience, satisfaction, and suggestions for improvement.

\subsection{Results}
\begin{figure}
    \centering
    \includegraphics[width=\linewidth]{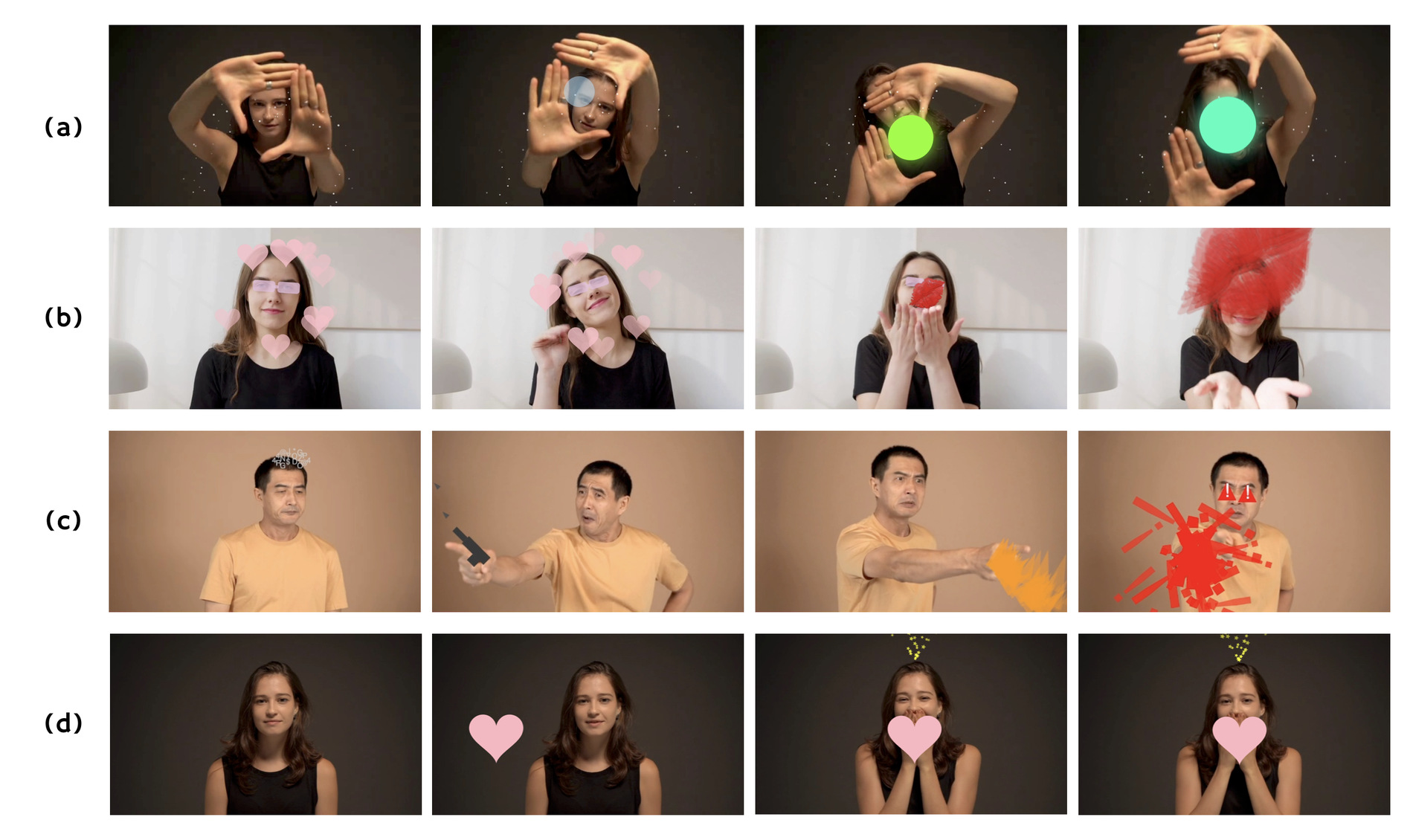}
    \caption{Representative results created by participants from the exploratory study.
    }
    \Description{Four images showing user-authored visual effects.
  (a) A girl controlling a scaling magic ball with her hand.
  (b) A girl with pink sunglasses and bubbles; a kiss graphic expands when she blows a kiss.
  (c) A man pointing in different directions, with distinct graphics emitting for each direction.
  (d) A heart moving toward a girl’s center, which then triggers a sparkle effect from the top of her head.}
    \label{fig:exploratory}
\end{figure}

\subsubsection{Qualitative Analysis}
Participants created 21 unique creative effects using PromptPoseFX, covering different aspects of the design space. Serval effects simultaneously utilized multiple categories within a single dimension through pose data. This was most prevalent in Dimension 2 (Feature Abstraction), where some effects paired Spatial coordinates with Geometric or Kinematic features(5/21). In these instances, the system resolved prompts into logic that used a joint’s position as a spatial anchor while concurrently using its speed as a kinematic driver or its relationship to other joints as a geometric vector. For example, Figure \ref{fig:exploratory}(b) shows a effect with pink sunglasses wore by the girl, where the position utilizes the eye position (Spatial Feature) and the rotation utilizes the angle of the line connecting two eyes (Geometric Feature).\\
We observed that the majority of the effects utilized a hybrid approach for controlling the effect timing, where they typically adopts an asynchronous duration or discrete time stamp to set the start and end of and effect, while allowing synchronous mapping between pose data and effect during the window. Notable, all participants utilizes the time marker provided by our system instead of explicitly mention the timing in prompt. U5 reflected that the time marker was very useful for creative exploration which allows them to freely adjust the timing without waiting for regeneration.

\subsubsection{Quantitative Analysis}
\begin{figure}
    \centering
    \includegraphics[width=\linewidth]{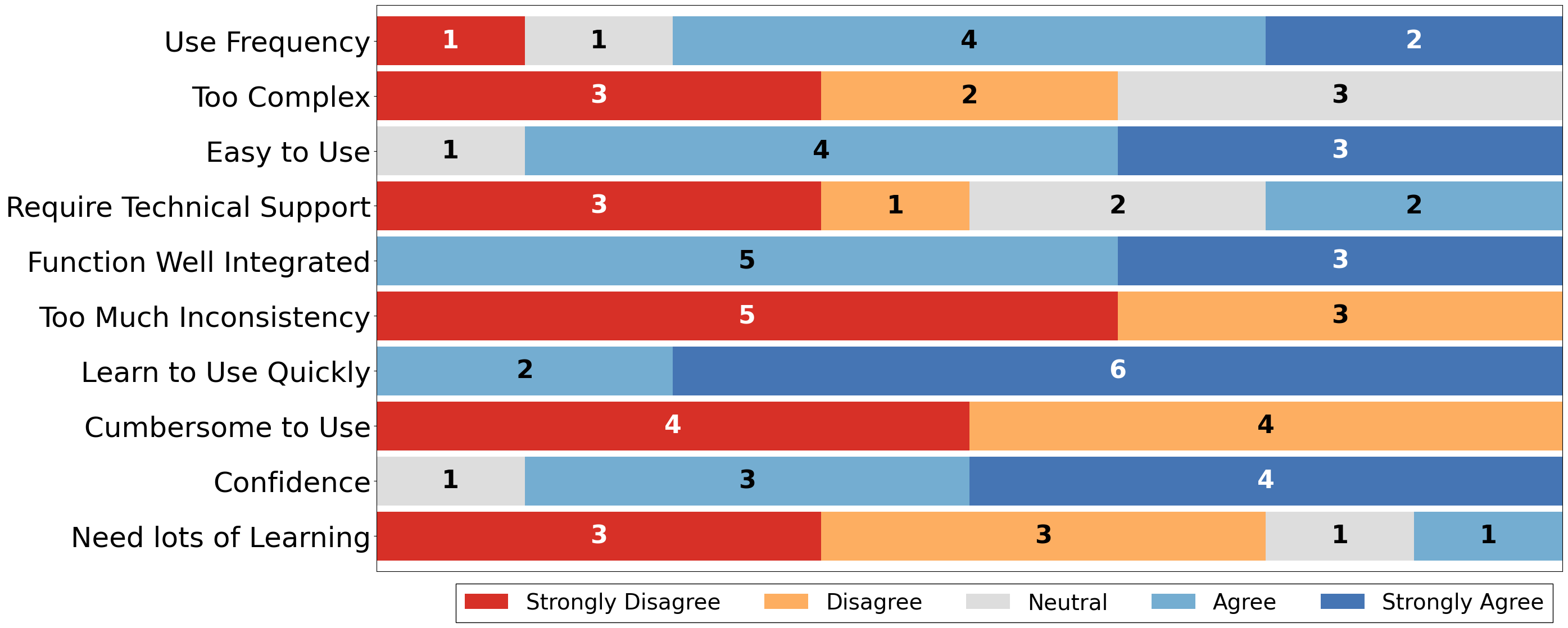}
    \caption{System Usability Scale response collected from participants of the exploratory study.}
    \Description{A horizontal stacked bar chart of the 10 SUS items. Odd-numbered positive items show a high concentration of 'Agree' and 'Strongly Agree' ratings. Even-numbered negative items show a high concentration of 'Strongly Disagree' and 'Disagree' ratings, indicating a high overall level of perceived usability and system consistency.}
    \label{fig:sus}
\end{figure}

\textbf{Usability} The mean SUS score was 80.62 (out of 100), indicating high usability. As shown in Figure \ref{fig:sus}, participants rated PromptPoseFX as easy to use (Mean = 4.25, SD = 0.71), easy to learn (Mean = 4.75, SD = 0.46), and fostering high user confidence (Mean = 4.38, SD = 0.74). In interviews, participants compared PromptPoseFX favorably to commercial tools: U2 noted it was "much simpler" than Adobe Premiere, which was "too complex and hard to get started with"; U4 observed it required less effort to memorize functionalities than CapCut and generated effects more quickly; U7 highlighted that it removed the "additional tracking step" required in CapCut for pose-driven effects.

\begin{table}[htbp]
\centering
\small
\begin{tabular}{lcc @{\hspace{2em}} lcc}
\toprule
\textbf{Dimension} & \textbf{Mean} & \textbf{SD} & \textbf{Dimension} & \textbf{Mean} & \textbf{SD} \\
\midrule
Exploration    & 17.25 & 1.83 & Enjoyment      & 18.0 & 2.07 \\
Expressiveness & 15.75 & 3.85 & Results        & 17.5 & 3.21 \\
Immersion      & 12.12 & 4.29 &                &      &      \\
\bottomrule
\end{tabular}
\caption{Creativity Support Index (CSI) Results (each dimension scored 0–20).}
\label{tab:csi_compact}
\end{table}
\textbf{Creativity Support} Participants gave high ratings across dimensions of exploration, enjoyment, and results worth effort. They especially valued real-time feedback and the ability to iterate quickly. Notably, U6 described their process as "a collaboration with the LLM"—when the system interpreted a vague prompt ("flying around the girl") in an unexpected but aesthetically pleasing way (placing butterflies around the head), they adopted this idea for refinement. U8 reported a similar experience where the system generated an additional "growing" animation on a "magic orb" effect, which they viewed as a "creative reminder" enabling rapid experimentation. These responses suggest LLMs' potential as proactive creative agents.

\subsubsection{Suggestions for Improvements}
Participants offered constructive feedback: U3 and U4 requested more complex, higher-fidelity effects; Some users suggested incorporating sketching for abstract effects difficult to describe with language; U6 proposed displaying visual examples related to user input to help refine language; U1 wished for support of additional human-related data (e.g., body contour, depth) to enable more varied effects.

\section{Discussion}
\subsection{Authoring Data-Driven Effects with LLMs}
PromptPoseFX introduces a novel mapping-based approach to create pose-driven visual effects, which reduces the need for manual keyframing. Our LLM pipeline automatically identifies relevant joints and defines the required mappings, such as velocity-to-scale and hand distance-to-opacity, based on inputs and generates corresponding procedural graphics scripts. Such design is generalizable beyond pose data for creating other dynamic data-driven visual effects. For example, with semantic object tracking data, our LLM pipeline can be prompted to create visual effects that react to object motion like basketball motion trail. By fine-tuning the system prompt to incorporate the target data form, our system can be adapted to other dynamic properties like audio frequencies, or environmental sensor data to create other data-driven visual effects. Our approach delegates the complex logic planning and implementation to LLMs, thus lowering the demand for manual operations and enabling creators to focus on high-level creative exploration.
\subsection{Interactive Control for Visual Content Generation}
While language-based interaction lowers the technical barrier for visual content creation, users often struggle to craft prompts that clearly reflect their intent \cite{interactionaugmented2025,bridgetg2024}. PromptPoseFX addresses this by providing interactive controls for prompting abstract concepts and refining generated results.\\
Our comparative study confirms that integrating UI-based controls into LLM systems enables users to specify intents with greater precision and less effort than text-only interaction. Furthermore, the generation of dynamic interfaces for post-generation refinement provides the fine-grained editing capabilities that were highly favored by participants.
\\
We believe that interactive control is essential for resolving language ambiguity in perception-heavy tasks like visual generation. We suggest that future systems integrate task-specific interfaces to help users express and refine their creation more efficiently.
\subsection{Effectiveness of Textual Feedback for Visual Content Generation}
Users may experience an “Intentionality Gap” when interacting with generative models, where they struggle to align their intention with the generated output\cite{bridgetg2024}. PromptPoseFX attempts to bridge this gap by providing textual feedback regarding the effect logic. Our user study results show that while visuals remain the primary editing reference, textual feedback can possibly help experienced users write editing commands more accurately. On the other hand, feedback from novice users pointed out that the current feedback language was too technical and hard to understand, indicating that transparency does not always equate to clarity. This suggests that different styles of textual feedback should be designed in future developments to accommodate users with different levels of expertise. In addition, as textual feedback is not commonly seen in traditional creative workflows, it also poses a challenge for integrating them naturally into existing creation processes.

\subsection{Limitation and Future work}
\textbf{Improve visual complexity and context-awareness}
The implementation of PromptPoseFX currently relies on p5.js for effect rendering and MediaPipe for pose tracking. These lightweight frameworks are good for fast 2D pose tracking and 2D effect rendering, but lack sufficient support to create complex visual effects with diverse video context. We observed from the user study that users sometimes used original video context as reference to describe the visual effects, such as prompting an object be scaled based on the actor’s head size. However, PromptPoseFX does not extract such information when processing the video, making such prompt invalid. We aim to refine the system to support a wider variety of human-related data, such as contours and depth, while integrating more powerful graphics frameworks to enable more diverse and complex visual effects.\\
\textbf{Integration of other input modalities}
While our current design received positive feedback from the users, it is still challenging to illustrate abstract concepts like specific shapes with the current system. For example, when creating the “heart” shape in the comparative study, some participants were not satisfied with the initial generated visuals and wanted to fine-tune the shape but struggled to describe the target shape through language. We envision that sketch inputs could help resolve this by enabling users to draw their desired visuals directly. We aim to investigate the needs of different modalities and incorporate them in future system development.

\section{Conclusion}
This paper introduced PromptPoseFX, an interactive system that leverages LLMs to create pose-driven visual effects. Our script-based approach supports non-destructive, independent effect generation and refinement. By automating the mapping from motion data to visual parameters, the system allows creators to focus on artistic design rather than technical details. To address the inherent ambiguity of natural language, we designed specific control mechanisms within a two-phase workflow. In the prompting phase, users can link text instructions to spatial coordinates and time markers, allowing precise effect placement. During the refinement phase, the system provides dedicated interfaces to adjust effect properties and timing without re-generating the results. Our user evaluation shows that these designs effectively balance automation with human control, significantly reducing the manual effort required to author complex visual effects.

\bibliographystyle{ACM-Reference-Format}
\bibliography{sample-base}

\appendix
\section{Analysis of Existing Tools}
Table \ref{tab:toolanalysis} provides a comparative summary of five video editing tools used for creating pose-driven visual effects. Regarding CapCut, our analysis focuses specifically on its pose-driven effect templates rather than its general video editing functionalities.
\begin{table*}
  \caption{Analysis of Video Editing Tools}
  \label{tab:toolanalysis}
  \begin{tabular}{p{2.5cm}p{2.5cm}p{2.5cm}p{2.5cm}p{2.5cm}p{2.5cm}}
    \toprule
     \textbf{Feature} & \textbf{Adobe Premiere Pro} & \textbf{Adobe After Effects} & \textbf{Final Cut Pro} & \textbf{DaVinci Resolve} & \textbf{CapCut} \\
    \midrule
    \textbf{Tracking Selection}
        & User-Initialized
        & User-Initialized
        & User-Initialized
        & User-Initialized
        & Template-defined \\
                \addlinespace

    \textbf{Pose Feature Extraction}
        & Not Supported
        & Scripting
        & Not Supported
        & Scripting
        & Not Supported\\
                \addlinespace

    \textbf{Data-to-Effect Mapping}
        & Manual Binding
        & Manual Binding \newline Script-based Binding
        & Manual Binding
        & Manual Binding \newline Script-based Binding
        & Template-defined \\
    \addlinespace
     \textbf{Graphic Asset Source}
        & Built-in Library\newline External Import \newline User Created
        & Built-in Library \newline External Import \newline User Created
        & Built-in Library \newline External Import
        & Built-in Library \newline External Import \newline User Created
        & Template-defined \\
        \addlinespace

    \textbf{Temporal Control}
        & UI Control
        & UI Control \newline Script Control
        & UI Control
        & UI Control \newline Script Control
        & UI Control\\
    \bottomrule
  \end{tabular}
\end{table*}

\section{Semantic Mapper Output Structure}
The followings are the output structure designed for the semantic mapper module.
For effect generation, the module takes in user input and outputs a specification with the following structure:\\
\begin{itemize}
    \item Control Target: The visual parameter to be modulated (e.g., circle position, star emission rate).
    \item Pose Data: The specific 2D pose feature used as input (e.g., right wrist position, left ankle moving speed).
    \item Description: A concise, plain-language explanation of how the pose data controls the control target
\end{itemize}
For effect modification, the module takes in user input and previously generated effect code for in-context control specification generation. The output structure is designed as follows:
\begin{itemize}
    \item Control Target: The specific parameter or effect component being changed
    \item Current Setting: Brief description of current implementation from code analysis
    \item Proposed Change: Description of requested change and its impact on pose data associations
\end{itemize}

\section{Code Generator Template}\label{codeTemplate}
The code template used for code generator is shown below:
{\small
\begin{verbatim}
export class effectClassName {
    constructor() { // Mandatory constructor
        // Initialize ALL effect parameters
    }
    update() { // Mandatory update method
        // Update time control
        // Update controllable parameters
        // Write effect rendering logic
    }
    // add self-defined helper functions below
}
\end{verbatim}
}
\section{Parameter Binder Output Example}\label{parameteroutput}
An example of the comment structure generated by parameter binder is shown below with the rendered UIs shown in Figure \ref{fig:systeminterface}:
{\small
\begin{verbatim}
/* PARAMETER-CONTROL
@effectName MusicSymbolsEffect
@param {number} MusicSymbolsEffect.trigger_distance:10 (0-50)
@param {number} MusicSymbolsEffect.note_count:20 (1-100) 
@param {number} MusicSymbolsEffect.note_lifetime:100 (1-300) 
@param {color} MusicSymbolsEffect.note_color:[255,255,0]       
@param {number} MusicSymbolsEffect.alpha:200 (0-255)      
*/
\end{verbatim}
}

\section{Survey Items} \label{surveyitem}
\subsection{Controllability}
The following items are used for evaluating spatial, temporal, and parameter control, as well as independent editing features, on a 5-point Likert scale as illustrated in Figure \ref{fig:likert}.
\begin{itemize}
    \item Controlling the timing of visual effects is easy.
    \item Controlling the spatial position of effects is easy.
    \item Adjusting effect parameters (e.g., color, stroke weight, size) is easy.
    \item My control over effect timing is precise.
    \item My control over effect positioning is precise.
    \item My control over parameter(e.g., color, stroke weight, size) values is precise.
    \item It is easy to edit individual effects without affecting others.
\end{itemize}
\subsection{Textual mapping guide}
The following statements are used for evaluating the user experience with the textual mapping guide, as illustrated in Figure \ref{fig:textguide}.
\begin{itemize}
    \item The text feedback helps me understand how the system generates and edits effects.
    \item The text feedback helps me write editing instructions more easily.
    \item The text feedback is aligned with the generated visual effect.
\end{itemize}

\end{document}